\documentclass[fontsize=11pt,a4paper]{scrartcl}

\makeatletter
\DeclareOldFontCommand{\rm}{\normalfont\rmfamily}{\mathrm}
\DeclareOldFontCommand{\sf}{\normalfont\sffamily}{\mathsf}
\DeclareOldFontCommand{\tt}{\normalfont\ttfamily}{\mathtt}
\DeclareOldFontCommand{\bf}{\normalfont\bfseries}{\mathbf}
\DeclareOldFontCommand{\it}{\normalfont\itshape}{\mathit}
\DeclareOldFontCommand{\sl}{\normalfont\slshape}{\@nomath\sl}
\DeclareOldFontCommand{\sc}{\normalfont\scshape}{\@nomath\sc}
\makeatother

\usepackage[T1]{fontenc}
\usepackage[utf8]{inputenc}
\usepackage[english]{babel}
\usepackage{graphicx}
\usepackage{authblk}
\usepackage[a4paper]{geometry}
\usepackage{lmodern}
\usepackage{color}

\usepackage{verbatim}
\usepackage{hyperref}
\usepackage{amsmath} 
\usepackage{amsfonts} 
\usepackage{amssymb}

\begin{document}

%
%
\title{Accretion outbursts in massive star formation}

%
%
\author[1]{D. M.-A.~Meyer}
\author[2,3]{E. I.~Vorobyov}
\author[1]{R.~Kuiper}
\author[1]{W.~Kley}
\affil[1]{\textcolor{black}{Institut} f. Astronomie und Astrophysik, Universit\" at T\" ubingen, Germany}
\affil[2]{Department of Astrophysics, The University of Vienna, Austria}
\affil[3]{Southern Federal University, Rostov-on-Don, Russia}
%
%
%

%
%
\date{}
\maketitle

%
%
\begin{abstract}
Using the HPC ressources of the state of Baden-Württemberg, we modelled \textcolor{black}{for the first time the} luminous burst from a young massive star by accretion of material 
from its close environment. We found that the surroundings of young massive stars \textcolor{black}{are} shaped as a clumpy disk whose fragments provoke outbursts once 
they fall onto the protostar and concluded that similar strong luminous events observed in high-mass star forming regions may be a signature of the presence 
of such disks.  
\end{abstract}
%
%

\begin{figure*}
        \centering
        \begin{minipage}[b]{ 0.24\textwidth}
                \includegraphics[width=1.0\textwidth]{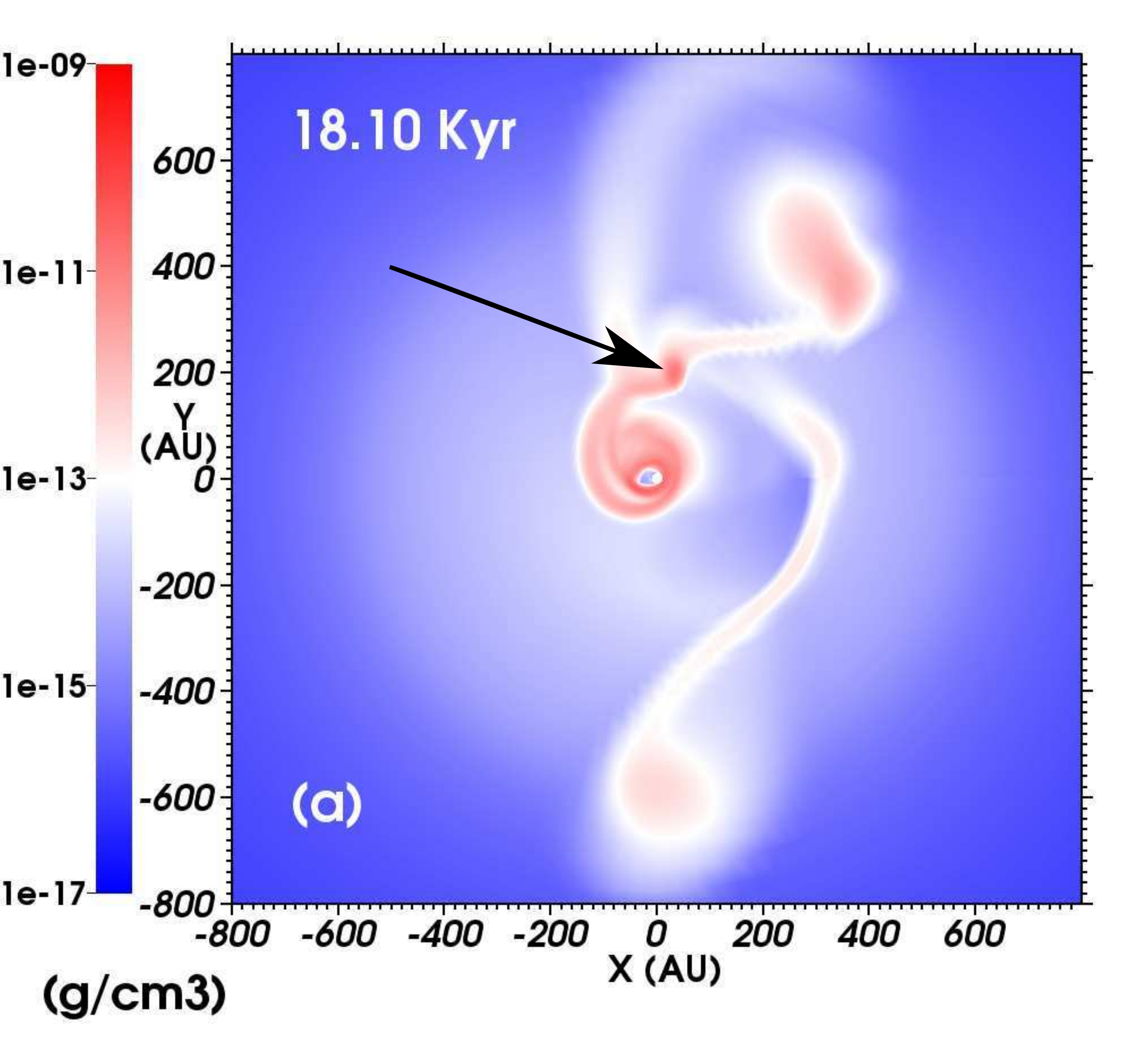}
        \end{minipage} 
        \begin{minipage}[b]{ 0.24\textwidth}
                \includegraphics[width=1.0\textwidth]{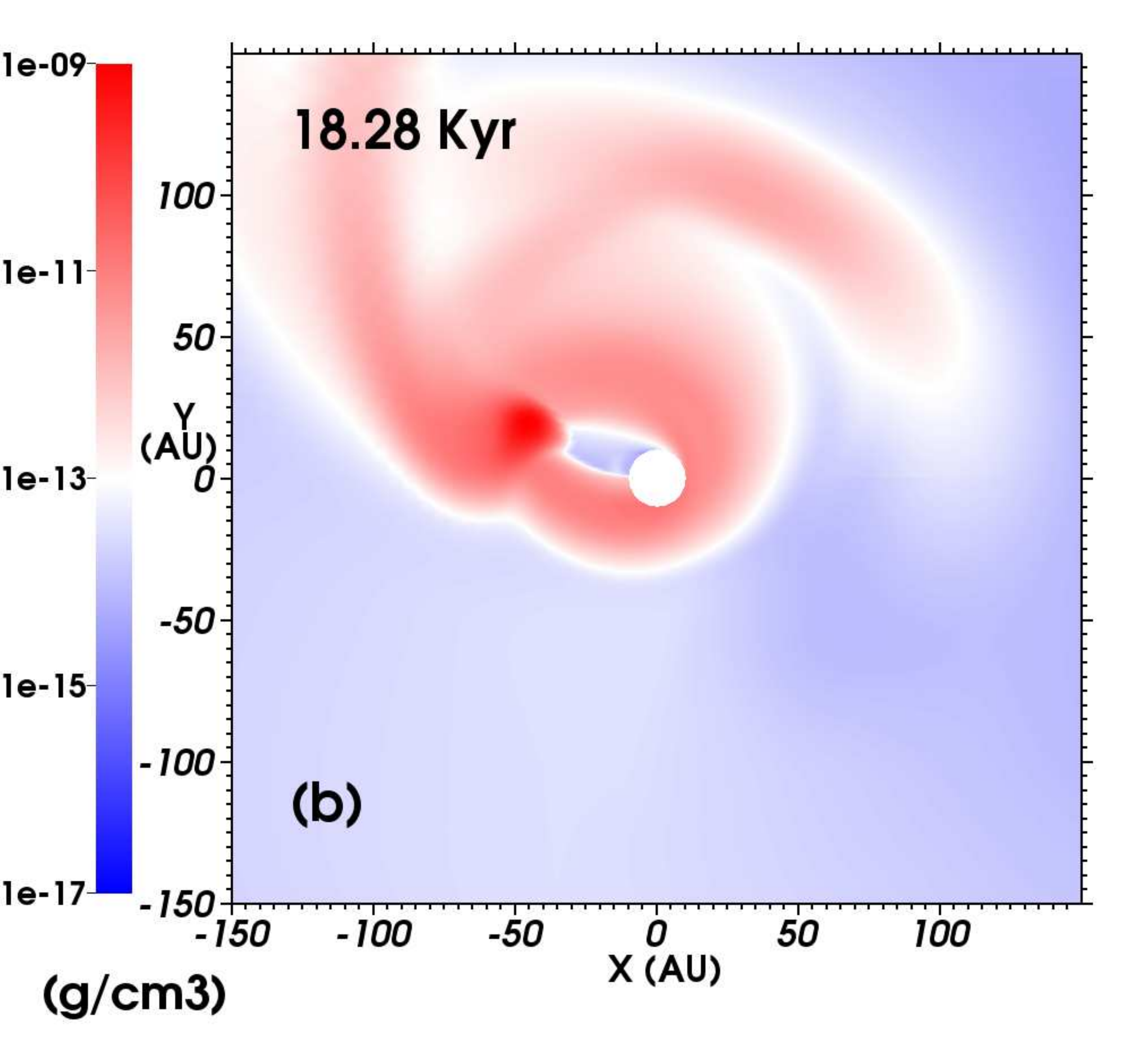}
        \end{minipage}  
        \begin{minipage}[b]{ 0.24\textwidth}
                \includegraphics[width=1.0\textwidth]{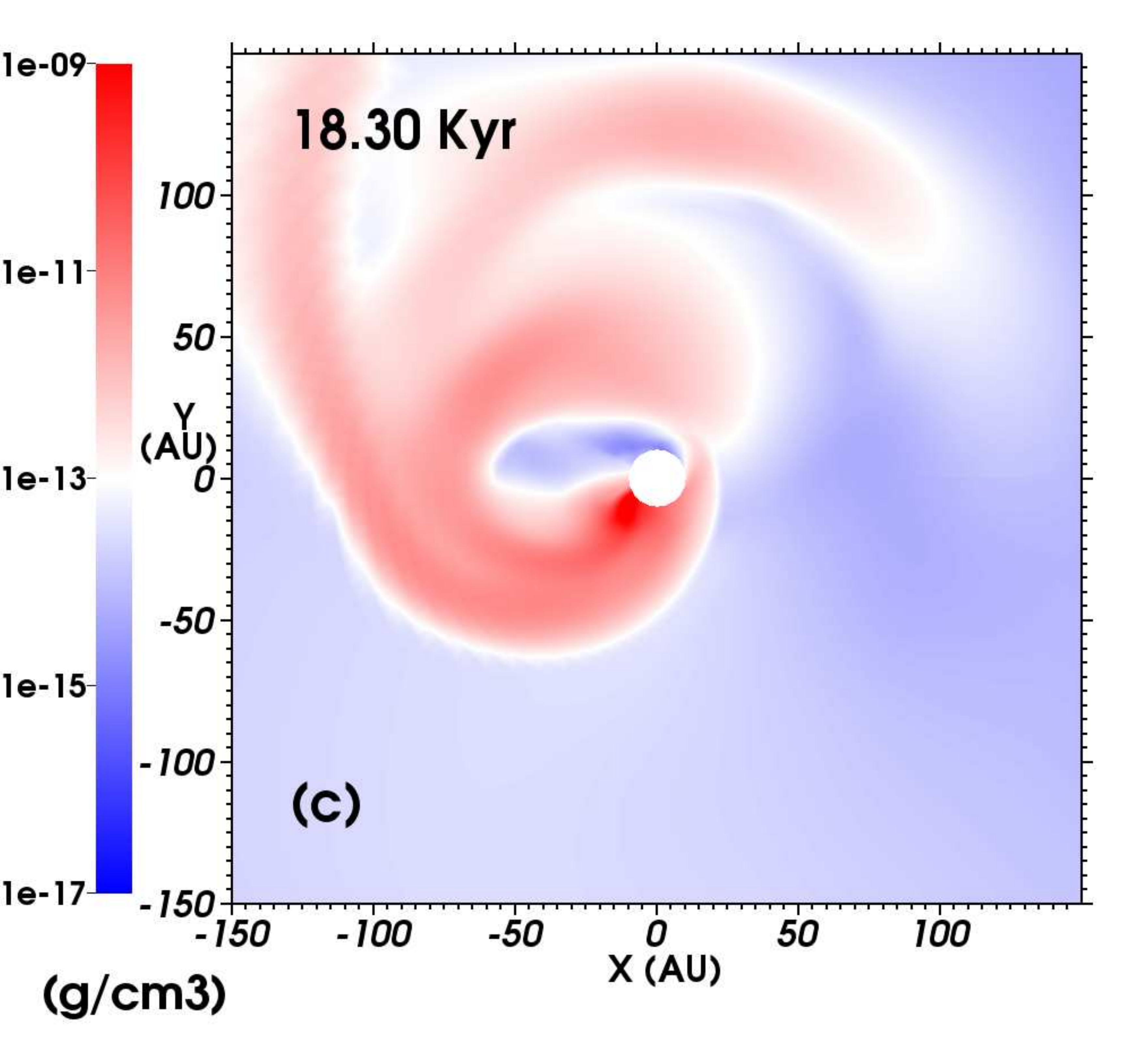}
        \end{minipage} 
        \begin{minipage}[b]{ 0.24\textwidth}
                \includegraphics[width=1.0\textwidth]{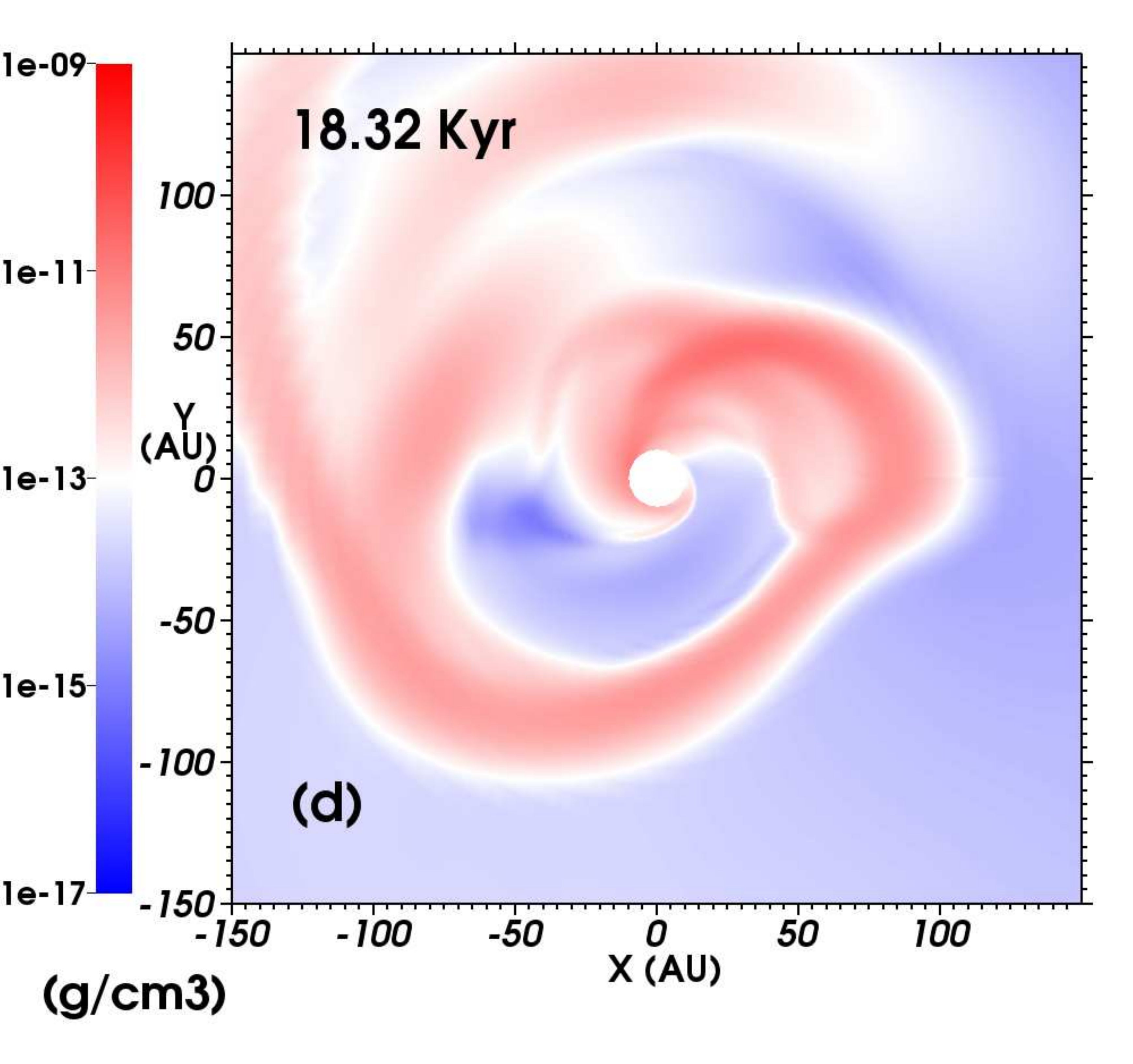}
        \end{minipage}
        \caption{ 
                 Midplane density in the center of the computational domain 
around the time of the outburst. (a) The region when a clump forms 
in a spiral arm. Panel (b-c) display zooms to 
illustrate the migration and accretion of a part of the clump. 
                 }      
        \label{fig:disk_density_plots}  
\end{figure*}

\section{An FU-Orionis-like burst from a high-mass protostar}
\label{sect:introduction}

\begin{figure}
        \centering              
        \begin{minipage}[b]{ 0.57\textwidth}
                        \includegraphics[width=0.98\textwidth]{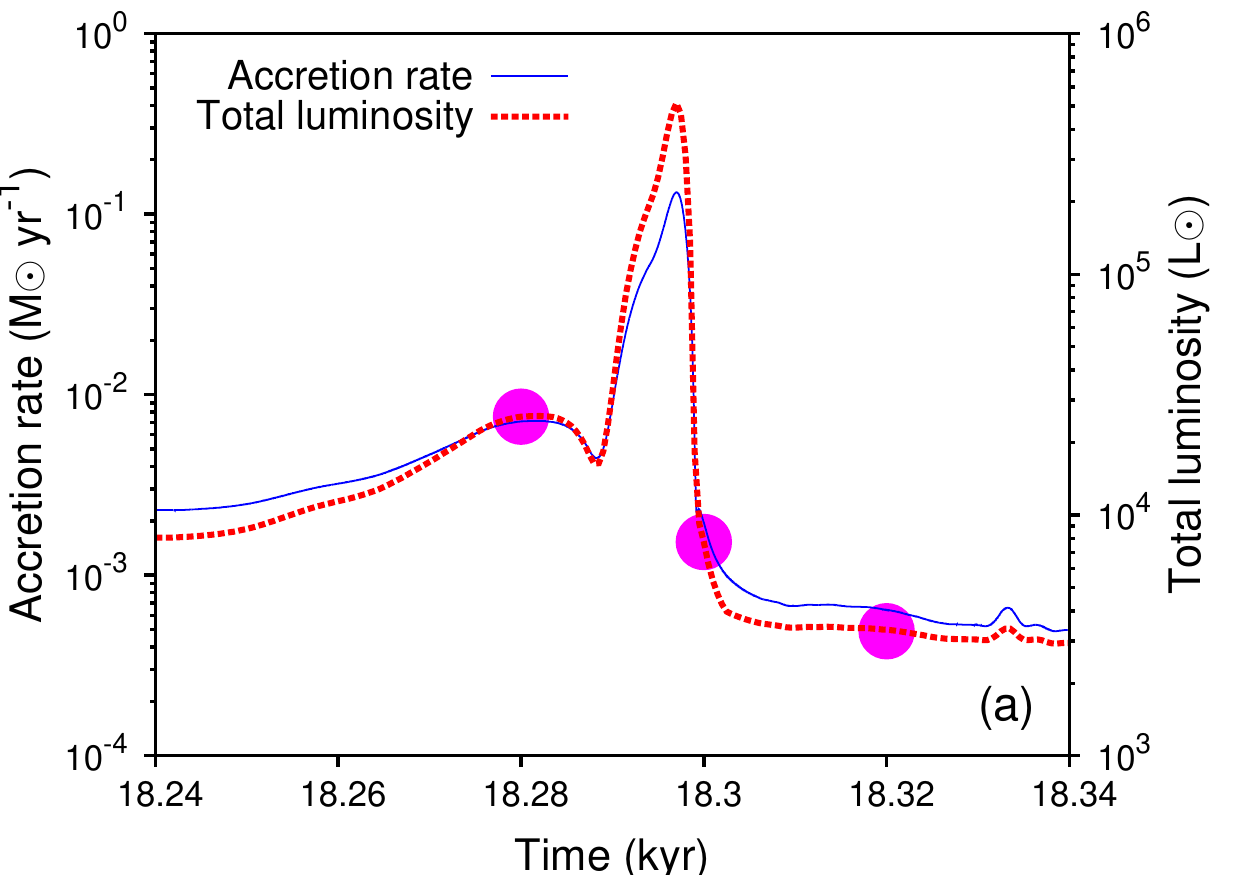}
        \end{minipage}         
        \caption{ 
                 Accretion rate onto the protostar (in $\rm M_{\odot}\, \rm yr^{-1}$) and total
luminosity of the protostar (in $\rm L_{\odot}$) during the burst. Magenta
dots mark the times of Fig.~\ref{fig:disk_density_plots}b-d.  
                 }      
        \label{fig:zoom_burst}  
\end{figure}

Accretion-driven luminosity outbursts are a vivid manifestation of variable mass 
accretion onto protostars. They are known as the so-called FU Orionis phenomenon 
in the context of low-mass protostars (Vorobyov \& Basu 2006). More recently, 
this process has been found in models of primordial star formation (Hosokawa et 
al. 2016). In Meyer et al. (2017), using numerical radiation hydrodynamics simulations of a 
collapsing $100\, \rm M_{\odot}$ pre-stellar cores rotating with a ratio of kinetic by 
gravitational energy of 4\% that produces a central massive protostar (cf. Kuiper et al. 2011), we stress 
that present-day forming massive stars also experience variable accretion (Fig.~\ref{fig:disk_density_plots}) and 
show that this process is accompanied by luminous outbursts induced by the 
episodic accretion of gaseous clumps migrating from the circumstellar disk onto 
the protostar (Fig.~\ref{fig:zoom_burst}).


\section{Observational implications}

We conjecture that luminous flares from regions hosting
forming high-mass star may be an observational
implication of the fragmentation of their accretion disks,
i.e. that those flares constitute a possible tracer of the
fragmentation of their accretion disks. This may apply to
the young star S255IR-NIRS3 that has recently been
associated to a 6.7 GHz methanol maser outburst
(Fujisawa et al. 2015, Stecklum et al. 2016) but also to the
other regions of high-mass star formation from which
originated similar flares (Menten et al. 1991) and which
are showing evidences of accretion flow associated to
massive protostars, see e.g. in W3(OH), W51 and W75.

\section*{Acknowledgements}

This study was conducted within the Emmy Noether research group on "Accretion Flows 
and Feedback in Realistic Models of Massive Star Formation" funded by the German Research 
Foundation under grant no. KU 2849/3-1. E.I.V. acknowledges support from the 
Austrian Science Fund (FWF) under research grant I2549-N27 and  RFBR grant 14-02-00719.

\end{document}